\documentclass[12pt]{article}

\usepackage{amssymb,amsfonts,amsmath,amscd}
\usepackage{graphicx}

\makeatletter \@addtoreset{equation}{section} \makeatother
\newtheorem{proposition}{Proposition}
\newtheorem{theorem}{Theorem}
\newtheorem{lemma}{Lemma}

\newcommand{\sumd}{\displaystyle\sum}

\begin{document}
\title{On universality of local edge regime for the deformed Gaussian
Unitary Ensemble}
\author{ T. Shcherbina\\
 Institute for Low Temperature Physics, \\
 47 Lenin Ave., 61103 Kharkov,
Ukraine. \\E-mail: t\underline{ }shcherbina@rambler.ru
}
\maketitle \centerline{{\it 2000 Mathematics Subject Classification.} Primary 15A52; Secondary 15A57}
\date{}

%
%
%
%
%

\maketitle

\begin{abstract}
We consider the deformed Gaussian ensemble $H_n=H_n^{(0)}+M_n$ in which $H_n^{(0)}$ is a hermitian matrix
(possibly random) and $M_n$ is the Gaussian unitary random matrix (GUE) independent of
$H_n^{(0)}$. Assuming that the Normalized Counting Measure of $H_n^{(0)}$ converges weakly (in probability
if random) to a non-random measure $N^{(0)}$ with a bounded support and assuming  some
conditions on the convergence rate, we prove
 universality of the local eigenvalue statistics near the edge of the limiting spectrum of $H_n$.
\end{abstract}

\section{Introduction}
Consider the deformed Gaussian Unitary Ensemble (DGUE)
\begin{equation}
H_n=H_n^{(0)}+M_n,  \label{H_n}
\end{equation}
where $H_n^{(0)}$ is a hermitian $n\times n$ matrix (possibly random, and in this case independent of $M_n$) with
eigenvalues $ \{h_{j}^{(n)}\}_{j=1}^{n}$ and $M_n$ is the Gaussian Unitary Ensemble matrix, defined as
\begin{equation}
M_n=n^{-1/2}W_n,  \label{M_n}
\end{equation}
where $W_n=\{W_{jk}\}_{j,k=1}^n$ is a hermitian $n\times n$ matrix whose entries $W_{jk}$ are independent
(modulo symmetry) Gaussian random variables such that
\begin{equation}
\mathbf{E}\{W_{jk}\}=\mathbf{E}\{W_{jk}^2\}=0,\quad \mathbf{E}\{|W_{jk}|^2\}=1, \quad j,k=1,..,n.  \label{W}
\end{equation}

Denote $\lambda_1^{(n)},\ldots,\lambda_n^{(n)}$ the eigenvalues of (\ref{H_n}). Define the Normalized Counting
Measure (NCM) of eigenvalues of the matrix as
\begin{equation}  \label{NCM}
N_n(\triangle)=\sharp\{\lambda_j^{(n)}\in \triangle,j=\overline{1,n} \}/n,\quad N_n(\mathbb{R})=1,
\end{equation}
where $\triangle$ is an arbitrary interval of the real axis.
Introduce also the NCM of eigenvalues of $H_n^{(0)}$
\begin{equation}\label{Nn0}
N_{n}^{(0)}(\triangle )=n^{-1}\sharp \{h_{j}^{(n)}\in \triangle ,j=\overline{1,n} \}.
\end{equation}
The behavior of $N_n$ as $n\to\infty$ is studied well enough. In particular, it was shown
in \cite{Pa:72} that if $N_{n}^{(0)}$ converges weakly, in probability if random, to a non-random
measure $N^{(0)}$ as $n\rightarrow \infty $, then $N_{n}$ also converges weakly in probability to a non-random measure
$N$, which is called the limiting NCM of the ensemble. The Stieltjes transforms $f$ of $N$ and $f^{(0)}$ of $ N^{(0)}$ are related as
\begin{equation}\label{Pe0}
f(z)=f^{(0)}(z+f(z)).
\end{equation}
Moreover, $N$ is absolutely continuous and its density $\rho$ is a bounded continuous function (see e.g.
\cite{TSh:09}).
These results characterize the so called global distribution of the eigenvalues of $H_n$.

The local regime deals with the behavior of eigenvalues of $n\times n$
random matrices on the intervals whose length is of the order of the mean distance between nearest eigenvalues.
According to the universality conjecture (see e.g. \cite{Me:91}, Chapter 19) the behavior
does not depend on the matrix probability
law (ensemble) and may only depend on the type of matrices (real
symmetric, hermitian, or quaternion real in the case of real eigenvalues and orthogonal,
unitary or symplectic in the case of the eigenvalues on the unit circle).
Usually two basic cases of universality are considered: universality in the bulk of the spectrum and universality
at the edges of the spectrum.
The local bulk regime, i.e. the distribution of eigenvalues near the points $\lambda$ in which the limiting
eigenvalue density $\rho(\lambda)\ne 0$, is studied for many ensembles of random matrices (see e.g. \cite{De-Co:99},
\cite{Pa-Sh:97,Pa-Sh:07}, \cite{TaoVu:09}, \cite{Er:10}). In particular,
universality for the DGUE (\ref{H_n}) was proved in \cite{Jo:01,Jo:09} for $H^{(0)}_n$ being the
Wigner matrix (i.e. the hermitian random matrix with i.i.d. (modulo symmetry) entries), in
\cite{Bl-Ku:04,Ap-Co:05} for $H^{(0)}_n$ being the matrix with only
two eigenvalues $\pm a$ of equal multiplicity, and in \cite{TSh:09} under the certain rather weak conditions both for random and
non-random $H^{(0)}_n$.
The local edge regime, which deals with the behavior of the eigenvalues near the edges of the spectrum (see
a definition below), is also studied for many ensembles of random matrices (see e.g. \cite{De-Co:99}, \cite{Pa-Sh:03},
\cite{Sosh:99}, \cite{Sosh:02}, \cite{BenPe:05}, \cite{TaoVu:09_ed}, \cite{Er:10}). In \cite{Jo:09} it was studied
for the special case of DGUE when $H_n^{(0)}=n^{-1/2}W^{(0)}$, where $W^{(0)}$
is a hermitian Wigner random matrix with the finite fourth moment, i.e. the matrix with i.i.d.
(modulo symmetry) entries such that
\begin{eqnarray} \label{Jcond}
\mathbf{E}\{W_{jk}^{(0)}\} &=&\mathbf{E}\{(W_{jk}^{(0)})^{2}\}=0,\;\\
\mathbf{E}
\{|W_{jk}^{(0)}|^{2}\}&=&1,\;\sup_{j,k}\mathbf{E}\{|W_{jk}^{(0)}|^{4}\}<\infty ,\quad j,k=1,..,n.
\notag
\end{eqnarray}
In this case every functionally independent entry of $H_n$ is the sum of the Gaussian random variable and
the independent random variable $W_{jk}^{(0)}$, i.e. is the Gaussian divisible random
variable according to \cite{BenPe:05}.

 The edge local regime of DGUE with $H^{(0)}_n$ being the matrix with only
two eigenvalues $\pm a$ of equal multiplicity was studied also in \cite{Bl-Ku:04,Ap-Co:05}.

In the present paper we prove universality of local edge regime for DGUE with $H_n^{(0)}$
satisfying rather weak conditions. Note that since the probability law of $M_n$ is unitary invariant, we can assume without loss of
generality that $H_n^{(0)}$ is diagonal.

Introduce the $m$-point correlation function $R_m^{(n)}$ by the equality:
\begin{multline}  \label{R}
\mathbf{E}\bigg\{ \sum_{j_{1}\neq ...\neq j_{m}}\varphi_m (\lambda_{j_{1}},\dots,\lambda_{j_{m}})\bigg\}\\
=\int \varphi _{m} (\lambda_{1},\dots,\lambda_{m})R_{m}^{(n)}(\lambda_{1},\dots,\lambda_{m})
d\lambda_{1},\dots,d\lambda_{m},
\end{multline}
where $\varphi_{m}: \mathbb{R}^{m}\rightarrow \mathbb{C}$ is bounded, sectionally continuous and symmetric in its
arguments and the summation is over all $m$-tuples of distinct integers $j_{1},\dots,j_{m}=\overline{1,n}$.
Here and below integrals without limits denote the integration over the whole real axis.

Let also
\begin{equation}
E_{n}(\triangle )=\mathbf{P}\{\lambda _{j}^{(n)}\not\in \triangle ,\,j= 1,..,n\}  \label{gapp}
\end{equation}
be the gap probability, and define for any sectionally continuous function $\varphi: \mathbb{R}\to [0,1]$ of a finite support
\begin{equation}\label{E_phi}
E_n[\varphi]=\mathbf{E}_n\bigg\{\prod\limits_{j=1}^n\left(1-\varphi(\lambda_j^{(n)})\right)\bigg\},
\end{equation}
where $\mathbf{E}_n$ denotes the expectation with respect
to the product measure of the probability law $\mathbf{P}^{(h)}_n$ of $H_n^{(0)}$ and the
Gaussian law $\mathbf{P}^{(g)}_n$ of $M_n$ of (\ref{M_n}). The functional $E_n[\varphi]$ of (\ref{E_phi})
is known as a generating functional
of the correlation functions, because its functional derivatives with respect to $\varphi$ give
the correlation functions (\ref{R}).

We will call the spectrum the support of $N$ and say that $\lambda_0$ is a right hand edge if
\begin{equation}\label{r_ep}
\begin{array}{c}
\rho(\lambda)>0,\quad \lambda\in [\lambda_0-\delta,\lambda_0)\\
\rho(\lambda)=0,\quad\lambda\in [\lambda_0,\lambda_0+\delta]
\end{array}
\end{equation}
for a sufficiently small $\delta$ (the left hand edge can be
defined similarly).

Introduce also
\begin{equation}  \label{A}
A(x,y)=\displaystyle\frac{Ai'(x)Ai(y)-Ai(x)Ai'(y)}{x-y},
\end{equation}
where $Ai(x)$ is the Airy function
\begin{equation}\label{Ai}
Ai(x)=\dfrac{1}{2\pi}\int\limits_Se^{is^3/3+isx}d\,s
\end{equation}
with $$S=\{z\in \mathbb{C}|\arg z=\pi/6\,\,\hbox{or}\,\arg z=5\pi/6\}.$$

We formulate now the main results of the paper
\begin{theorem}
\label{thm:1} Let $H_n^{(0)}$ in (\ref{H_n}) be non-random and such that its Normalized Counting Measure
(\ref{Nn0}) converges weakly to a measure $N^{(0)}$ of a bounded support and let $\lambda_{0}$ be
a right hand edge of $\,\mathrm{supp}\,N$, where $N$ is the limiting NCM of (\ref{H_n}).
Denote $f$ the Stieltjes transform of $N$ and set
\begin{equation}\label{z_0}
z_0=\lambda_0+f(\lambda_0+i0).
\end{equation}
(it was proved in \cite{TSh:09} that there exists $\lim\limits_{\varepsilon\to+0}f(\lambda+i\varepsilon)$).
Assume also that

 (i) for any compact set $K\subset \mathbb{C}$ such that $\mathrm{dist}\,(K,\mathrm{supp}\,N^{(0)})>0$ we have
\begin{equation}\label{alpha}
\max\limits_{z\in K}\left|f_n^{(0)}(z)-f^{(0)}(z)\right|\le Cn^{-2/3-\alpha},\quad\alpha>0,
\end{equation}
where $C$ is independent of $n$.

 (ii)~$\mathrm{dist}\,(z_0,\mathrm{supp}\,N^{(0)})=d>0.$

 (iii)~$\lim\limits_{n\to\infty}\max\limits_{j=1,..,n}
 \mathrm{dist}\,(h_j^{(n)},\mathrm{supp}\,N^{(0)})=0$.\\
 Then we have:

(1) for
\begin{equation}\label{gam}
\gamma=\left(\displaystyle\int\dfrac{N^{(0)}(d\,h)}{(z_0-h)^3}\right)^{-1/2}
\end{equation}
and any fixed $m$ uniformly in $\xi_1, \xi_2,\ldots, \xi_m$ varying in any compact set in $\mathbb{R}$
\begin{multline}  \label{Un}
\lim\limits_{n\to \infty}\displaystyle\frac{1}{(\gamma n)^{2m/3}}
R^{(n)}_m\left(\lambda_0+\displaystyle\frac{\xi_1}{(\gamma n)^{2/3}},
\ldots,\lambda_0+\displaystyle\frac{\xi_m}{(\gamma n)^{2/3}}\right)\\
=\det \{A(\xi_i,\xi_j)\}_{i,j=1}^m,
\end{multline}
where $R^{(n)}_m$ and $A$ is defined in (\ref{R}) and (\ref{A}) respectively.

(2) for $\Delta=[a,b]\subset \mathbb{R}$ with $n$-independent $a$ and $b$ and $\Delta_n=\lambda_0+
\Delta/(\gamma n)^{2/3}$ there exists a limit of the gap probability (\ref{gapp})
\begin{equation}\label{gp}
\lim_{n\rightarrow \infty }E_{n}(\Delta_{n})=\det(1-A_\Delta)
\end{equation}%
i.e., the limit is the Fredholm determinant of the integral operator $A_\Delta$, defined in
$L_2(\Delta)$ by the kernel (\ref{A}). The same formula is valid for $\Delta_n=[\lambda_0+a/(\gamma n)^{2/3}, b]$
with $n$-independent $b$ or $\Delta_n=[\lambda_0+a/(\gamma n)^{2/3}, \infty]$, if $\Delta_n$ does not contain the edges
of $\,\mathrm{supp}\, N$ except may be $\lambda_0$.
\end{theorem}

\textit{Remarks}

1. For the left hand edges the statement is similar.

2. Note that for many known ensembles of random matrices $\alpha=1/3$ (see e.g. \cite{Jo:98},\cite{Lyt-Pa:09}).

3. A sufficient condition to have condition (ii) of Theorem \ref{thm:1} is

(ii') for any $\lambda$ which is an edge of $\mathrm{supp}\,N^{(0)}$  we have
\[
\displaystyle\int\dfrac{N^{(0)}(d\,h)}{(h-\lambda)^2}>1.
\]
Indeed, it follows from \cite{TSh:09} that
\[
\displaystyle\int\dfrac{N^{(0)}(d\,h)}{(h-\lambda_0-f(\lambda_0+i0))^2}\le 1.
\]
Hence, (ii') implies $\lambda_0+f(\lambda_0+i0)$ is not an edge of $\mathrm{supp}[N^{(0)}]$
and\\
 $\lambda+f(\lambda+i0)\not\in \mathrm{supp}\,N^{(0)}$.

4. It will be proved below (see Proposition \ref{p:lim_cont} and Remark 3 of Section 2) that under the conditions
of Theorem \ref{thm:1}
we have $$\displaystyle\int\dfrac{N^{(0)}(d\,h)}{(z_0-h)^3}>0$$
and
\begin{equation}\label{gamma}
\rho(x)=\dfrac{\gamma}{\pi}\sqrt{|x-\lambda_0|}(1+o(1)),\quad x\to\lambda_0-0.
\end{equation}

\begin{theorem}
\label{thm:2} Let the eigenvalues $\{h_{j}^{(n)}\}_{j=1}^{n}$ of $H_n^{(0)} $ in (\ref{H_n}) be random variables independent of $W_n$ of (\ref{W}) and let $\lambda_{0}$ be
a right hand edge of $\mathrm{supp}\,N$, where $N$ is the limiting NCM of (\ref{H_n}). Assume that

(i) there exists a non-random measure
$N^{(0)}$ of a bounded support such that for the Stieltjes transforms $f^{(0)}$ of $N^{(0)}$ and $g^{(0)}_n$
of $N^{(0)}_n$ and for any compact set $K\subset \mathbb{C}$ such that \\
$\mathrm{dist}\,(K,\mathrm{supp}\,N^{(0)})>0$ we have
\begin{equation}\label{conpN0}
\lim\limits_{n\to\infty}{\bf P}^{(h)}_n\{|g_n^{(0)}(z)-f^{(0)}(z)|>n^{-2/3-\alpha}\}=0,\quad\alpha>0
\end{equation}
uniformly in $z\in K$. Here and below ${\bf P}^{(h)}_n\{\ldots\}$ denotes the probability law of $\{h_j^{(n)}\}_{j=1}^n$.

(ii) $\mathrm{dist}\,(z_0,\mathrm{supp}\,N^{(0)})=d>0$, where $z_0$ is defined in (\ref{z_0}).

(iii) for any $\delta>0$
\[\lim\limits_{n\to\infty}{\bf P}^{(h)}_n\{\exists j\in\{1,..,n\}:
\mathrm{dist}\,(h_j^{(n)},\mathrm{supp}\,N^{(0)})>\delta\}=0.
\]
Then for any sectionally continuous function $\varphi: \mathbb{R}\to[0,1]$ of a finite support we have
\begin{equation}\label{gp_sl}
E[\varphi]:=\lim\limits_{n\to\infty}E[\varphi_n]=
\det\left(1-\varphi^{1/2}A\varphi^{1/2}\right),
\end{equation}
where $\varphi_n(x)=\varphi(n^{2/3}\gamma^{2/3}(x-\lambda_0))$ and
$E_n[\varphi]$ is defined in (\ref{E_phi}). Here
the r.h.s. is the Fredholm determinant on $L^2(\mathbb{R})$
with kernel $\varphi^{1/2}A\varphi^{1/2}$, where $A$ is defined in (\ref{A}).
\end{theorem}
\textit{Remarks}

1. If $\varphi=\chi_\Delta$, where $\Delta$ is the same as in Theorem \ref{thm:1}, then (\ref{gp_sl})
implies (\ref{gp}). The universal form of (\ref{gp_sl}) is one of possible
(although more weak than (\ref{Un})) forms of universality of correlation functions.

2. The conditions of Theorem \ref{thm:2} hold for the case, where
$H_n^{(0)}=n^{-1/2}W^{(0)}$ is the Wigner matrix, satisfying (\ref{Jcond}), considered in \cite{Jo:09}.
Indeed, the condition (i) in this case follows from the Chebyshev inequality and the bounds
(see e.g. \cite{Lyt-Pa:09})
\[
\mathbf{E}^{(h)}_n\{|g_n^{(0)}(z)-f_n^{(0)}(z)|^2\}\le Cn^{-2},\quad |f_n^{(0)}(z)-f^{(0)}(z)|\le Cn^{-1}
\]
valid uniformly in $z\in K$, where $K\subset \mathbb{C}$ is a compact set such that \\
$\mathrm{dist}\,(K,\mathrm{supp}\,N^{(0)})>0$,
$\mathbf{E}^{(h)}_n$ denotes the expectation with respect
to the measure generated by $H_{n}^{(0)}$, $g_n^{(0)}$ and $f_n^{(0)}$ are the Stieltjes transforms
of $N_n$ and $\,\mathbf{E}^{(h)}_n\{N_n\}$ respectively.
Conditions (ii) of Theorem \ref{thm:2} for the Wigner Ensembles  can also be easily checked, because equation (\ref{Pe0}) for
$f$ is quadratic. The result \cite{Ba-Yi:88} yields (iii).

The paper is organized as follows. In Section $2$ we prove Theorem \ref{thm:1} using an extension of the techniques in
\cite{TSh:09}. The techniques are based on the steepest descent method applied to the determinant formulas for
the correlation functions (\ref{R}),
which were obtained in \cite{Br-Hi:97,Jo:01,TSh:09}. Section $3$
deals with the proof of auxiliary statements for Theorem~\ref{thm:1}. Theorem~\ref{thm:2} is proved
in Section~$4$.

We denote by $M, C, C_1$, etc. various constants appearing below, which
can be different in different formulas, but are independent of $n$. We denote also
 $U_\delta(a)=(a-\delta,a+\delta)$.

\section{The proof of Theorem \ref{thm:1}}
To prove Theorem \ref{thm:1} we need the determinant formulas for the correlation functions (\ref{R}),
which were obtained in \cite{Br-Hi:97,Jo:01,TSh:09}.
\begin{proposition}\label{p:det_f}
Let $H_n$ be the random matrix (\ref{H_n}) and $\{R_m^{(n)}\}_{m=1}^n$ be the correlation functions (\ref{R})
of its eigenvalues.
Then we have for every $m=1,..,n$
\begin{equation}  \label{Det}
R_m^{(n)}(\lambda_1,\ldots,\lambda_m)=\det\{K_n(\lambda_i,\lambda_j)\}
\end{equation}
with
\begin{equation}  \label{K}
K_n(\lambda,\mu)
=-n\displaystyle\int\limits_{l}\displaystyle\frac{d\,t}{2\pi}\oint\limits_{L} \displaystyle\frac{d\,v}{2\pi}
\displaystyle\frac{e^{ -\frac{n}{2}(v^2-2v\lambda-t^2+2\mu\,t))}
}{v-t}\prod\limits_{j=1}^n \left(\displaystyle\frac{t-h_j^{(n)}} {v-h_j^{(n)}}\right),
\end{equation}
where $l$ is a line parallel to the imaginary axis and lying to the left of all $\{h_j^{(n)}\}_{j=1}^n$, and
$L$ is a closed contour, encircling $ \{h_j^{(n)}\}_{j=1}^n$ and not intersecting $l$.
\end{proposition}
Set
\begin{equation}\label{K_kal}
\mathcal{K}_n(\xi,\eta)=n^{-2/3}
K_n\left(\lambda_0+\dfrac{\xi}{ n^{2/3}},\lambda_0+\dfrac{\eta}{ n^{2/3}}\right).
\end{equation}
In view of (\ref{Det}) (\ref{Un}) follows from the relation
\begin{equation}\label{limK}
\lim\limits_{n\to\infty}\gamma^{-2/3}\theta(\xi,\eta)\mathcal{K}_n(\xi/\gamma^{2/3},\eta/\gamma^{2/3})=
A(\xi,\eta),
\end{equation}
where $\xi,\eta\in \mathbb{R}$, $|\xi|,|\eta|<M<\infty$, $\gamma$ and $A$ are defined in
(\ref{gam}) and (\ref{A}) respectively, and $\theta(\xi,\eta)$ is any function such that
\begin{multline}\label{theta}
\det\left\{\theta(\xi_i,\xi_j)
\mathcal{K}_n(\xi_i/\gamma^{2/3},\xi_j/\gamma^{2/3})\right\}_{i,j=1}^m\\
=\det\left\{\mathcal{K}_n(\xi_i/\gamma^{2/3},\xi_j/\gamma^{2/3})\right\}_{i,j=1}^m.
\end{multline}
Putting in (\ref{K}) $ \lambda=\lambda_0+\xi/n^{2/3}$ and $\mu=\lambda_0+\eta/n^{2/3}$, we get
\begin{multline}  \label{Ker}
\mathcal{K}_n(\xi,\eta)=-n^{1/3}
\displaystyle\int\limits_l\displaystyle\frac{dt}{2\pi} \oint\limits_{L}\displaystyle\frac{dv}{2\pi}
\exp\{n^{1/3}(v\xi-t\eta)\}\\
\displaystyle\frac{\exp\{n(S_n(t,\lambda_0)- S_n(v,\lambda_0))\}}{v-t },
\end{multline}
where
\begin{equation}  \label{S_n}
S_n(z,\lambda)=\displaystyle\frac{z^2}{2}+\displaystyle\frac{1}{n} \sum\limits_{i=1}^{n}\log(z-h_j^{(n)})-
\lambda z-S^*
\end{equation}
with a constant $S^*$ which will be chosen later (see (\ref{S*})). Here $L$ and $l$ are as in the
Proposition \ref{p:det_f}.

Let us choose a contour $L$ in (\ref{Ker}) as a special $n$-dependent contour that will be denoted $L_n$.
To describe it consider
\begin{equation}  \label{f_0,n}
f^{(0)}_n(z)=\displaystyle\frac{1}{n}\displaystyle\sum\limits_{j=1}^n \displaystyle\frac{1}{h_j^{(n)}-z},
\end{equation}
and the equation
\begin{equation}  \label{eqv_f_0,n}
z-f^{(0)}_n(z)=\lambda
\end{equation}
for given $\lambda\in \mathbb{R}$. The equation is a polynomial equation of degree $(n+1)$ in $z$, hence
it has $(n+1)$ roots. Since the l.h.s. of
(\ref{eqv_f_0,n}) tends to $+\infty$, if $z\in\mathbb{R}\to h_j^{(n)}+0$, and the l.h.s. tends to $-\infty$, if
$z\in\mathbb{R}\to h_j^{(n)}-0$, the $n-1$ roots are always real and belong to the segments between adjacent
$h_j^{(n)}$'s . If $\lambda$ is big enough, then all $n+1$ roots are real. Let $z_n(\lambda)$ be a real root
equal to $\lambda-1/\lambda+O(1/\lambda^2) $, as $\lambda\to\infty$. If $\lambda$ decreases, then
$z_n(\lambda)$ decreases too, and coming to some $\lambda_{c_1}$ the real root disappears and there appear
two complex ones: $z_n(\lambda)$ and $\overline{ z_n(\lambda)}$. Then $z_n(\lambda)$ may be real again,
then again complex, and so on, however as soon as $\lambda$ becomes less then some $\lambda_{c_2} $, the root
becomes real again. We set
\begin{multline}  \label{L_n}
L_n=\{z\in\mathbb{C}:z=z_n(\lambda),\,\Im z_n(\lambda)>0\}
\cup\\
\{z\in\mathbb{C}:z=\overline{z_n(\lambda)},\,\Im z_n(\lambda)>0\}\cup S,
\end{multline}
where $S$ is a set of points $z=z_n(\lambda)$ in which $z_n(\lambda)$ becomes real.
It is clear that the set of corresponding $\lambda$'s is $\bigcup\limits_{j=1}^k I_k$, where
$\{I_j\}_{j=1}^k$ are non intersecting segments, and that $L_n$ is closed and encircles
$\{h_j^{(n)}\}_{j=1}^n$.

Let us consider the limiting equation
\begin{equation}  \label{eqv_f_0}
V(z):=z-f^{(0)}(z)=\lambda, \quad
\end{equation}
where $\lambda\in \mathbb{R}$ is fixed and $f^{(0)}$ is the Stieltjes transform of the limiting NCM $N^{(0)}$ of
$H_n^{(0)}$. We have

\begin{proposition}\label{p:lim_cont}
Under conditions of Theorem \ref{thm:1} the limiting measure $N$ is
absolutely continuous and its density $\rho$ is continuous. Moreover, equation (\ref{eqv_f_0}) for $\lambda=\lambda_0$
has a unique solution $z_0$ of (\ref{z_0}) of the multiplicity two.
The solution is real and satisfies the relations
\begin{equation}\label{eqv_z_0}
\displaystyle\int\dfrac{N^{(0)}(d\,h)}{(z_0-h)^2}=1,\quad
\displaystyle\int\dfrac{N^{(0)}(d\,h)}{(z_0-h)^3}>0,
\end{equation}
\end{proposition}
and also
\begin{lemma}\label{l:z*_0,n}
There exists $n_0$ such that if $n>n_0$, then
\begin{equation}\label{eqv_f,pr}
\dfrac{d}{d\,z}f^{(0)}_n(z)=1,\quad |z-z_0|\le \delta,
\end{equation}
has a unique solution $z^*_{0,n}$ for any sufficiently small $\delta$, and the solution satisfies the
inequality
\begin{equation}\label{*}
|z^*_{0,n}-z_0|\le n^{-1/3-\varepsilon}
\end{equation}
for some $\varepsilon>0$, where $z_0$ is defined in (\ref{z_0}), and
\begin{equation}\label{pol_ga_n}
\dfrac{d^2}{d\,z^2}f^{(0)}_n(z_{0,n}^*)<-C<0.
\end{equation}
Moreover, we have
\begin{equation}\label{z-z_0}
|z_0-z_n(\lambda_0)|\le n^{-1/3-\varepsilon}
\end{equation}
for some $\varepsilon>0$, where $z_n(\lambda)$ is a solution of (\ref{eqv_f_0,n}) such that
$z_n(\lambda)=\lambda-1/\lambda+O(1/\lambda^2) $, as $\lambda\to\infty$
\end{lemma}
The proofs of Proposition \ref{p:lim_cont} and Lemma \ref{l:z*_0,n} are given in the next Section.
Set
\begin{equation}\label{lam_0,n}
\lambda_{0,n}=z^*_{0,n}-f^{(0)}_n(z^*_{0,n})
\end{equation}
and choose $S^*$ in (\ref{S_n}) as
\begin{equation}\label{S*}
S^*=(z^*_{0,n})^2/2+ \dfrac{1}{n}\sum\limits_{j=1}^n\log (z^*_{0,n}-h_j^{(n)})-\lambda_{0,n}z^*_{0,n}.
\end{equation}
Then (\ref{Ker}) can be rewritten as
\begin{eqnarray}  \label{Ker_osn1}
\mathcal{K}_n(\xi,\eta)&=&-n^{1/3}
\displaystyle\int\limits_l\displaystyle\frac{dt}{2\pi} \oint\limits_{L_n}\displaystyle\frac{dv}{2\pi}
e^{n^{1/3}(v\xi-t\eta)+n(\lambda_{0,n}-\lambda_0)(t-v)}\\ \notag
&&\times\displaystyle\frac{\exp\{n(S_n(t,\lambda_{0,n})- S_n(v,\lambda_{0,n}))\}}{v-t },
\end{eqnarray}
where $L_n$ is defined in (\ref{L_n}) and $l$ is a line parallel to the imaginary axis and lying to the left
of $L_n$.

 The next step is to replace $l$ in (\ref{Ker}) by
\begin{equation}\label{l_n}
l_n=\{z\in\mathbb{C}:z=\zeta_n(y)=z^*_{0,n}+i\,y,\,\, y\in\mathbb{R}\}.
\end{equation}
We are going to use the steepest descent method, i.e. to show that only integrals in a small neighborhood of $z_{0,n}^*$
give the non vanishing contribution in the r.h.s. of (\ref{Ker_osn1}). This requires the knowledge of
the behavior of $\Re S_n(z,\lambda_{0,n})$ on $L_n$ of (\ref{L_n}) and
$l_n$ of (\ref{l_n}).
\begin{lemma}
\label{l:min_max}  The function $\Re\,S_n(z_n(\lambda),\lambda_{0,n})$ is monotone increasing for \\
$\lambda>\lambda_{0,n}$
and monotone decreasing for $\lambda<\lambda_{0,n}$, thus $\Re\,S_n(z,\lambda_{0,n})\ge 0$ for \\
$z\in L_n$, and
the equality holds only at $z=z^*_{0,n}$. Besides,
\begin{equation}\label{fn0<1}
\Re z^\prime_n(\lambda)=\Re \left(1- \dfrac{d}{d\,z}f^{(0)}_n(z_n(\lambda))\right)^{-1}>0
\end{equation}
for all $\lambda\in \mathbb{R}$.
Moreover, the function $\Re\,S_n(\zeta_n(y),\lambda_{0,n})$ with $\zeta_n(y)$ of (\ref{l_n})
is monotone increasing for $y<0$
and monotone decreasing for $y>0$, thus $\Re\,S_n(z,\lambda_{0,n})\le 0$ for $z\in l_n$, and
the equality holds only at $z=z^*_{0,n}$.
\end{lemma}
The proof of the lemma can be found in \cite{TSh:09}.
The lemma yields
\begin{equation}  \label{Main1}
\Re(n(S_n(t,\lambda_{0,n})-S_n(v,\lambda_{0,n})))\le 0,\quad t\in l_n,\,\,v\in L_n
\end{equation}
and the equality holds only if $v=t=z^*_{0,n}$.

Prove now that for $n>n_0$
\begin{equation}\label{razn_lam}
|\lambda_0-\lambda_{0,n}|\le C n^{-2/3-2\varepsilon}
\end{equation}
with $\varepsilon$ from Lemma \ref{l:z*_0,n} and $\lambda_{0,n}$ of (\ref{lam_0,n}). Indeed, using (\ref{eqv_f_0,n})
for $\lambda=\lambda_0$,
(\ref{lam_0,n}), and (\ref{eqv_f,pr}) we have
\begin{equation}\label{razn_lam1}
\begin{array}{c}
|\lambda_{0,n}-\lambda_0|
=|z_n(\lambda_0)-z^*_{0,n}|\cdot\left|1-\dfrac{1}{n}\sum\limits_{j=1}^n\dfrac{1}
{(z_n(\lambda_0)-h_j^{(n)})(z^*_{0,n}-h_j^{(n)})}\right|\\
=|z_n(\lambda_0)-z^*_{0,n}|\cdot\left|\dfrac{1}{n}\sum\limits_{j=1}^n\dfrac{1}
{(z^*_{0,n}-h_j^{(n)})^2}\right.\\ \left.-\dfrac{1}{n}\sum\limits_{j=1}^n\dfrac{1}
{(z_n(\lambda_0)-h_j^{(n)})(z^*_{0,n}-h_j^{(n)})}\right|\\
=|z_n(\lambda_0)-z^*_{0,n}|^2\cdot\left|\dfrac{1}{n}\sum\limits_{j=1}^n\dfrac{1}
{(z_n(\lambda_0)-h_j^{(n)})(z^*_{0,n}-h_j^{(n)})^2}\right|.
\end{array}
\end{equation}
Since $z^*_{0,n},z_n(\lambda_0)\in \omega_n$ (see Lemma \ref{l:z*_0,n}), we obtain
\[
|z_n(\lambda_0)-z^*_{0,n}|\le 2 n^{-1/3-\varepsilon}.
\]
Moreover, taking into account conditions (ii) -- (iii) of Theorem \ref{thm:1}, we get for $n>n_0$
\[
|z_n(\lambda_0)-h_j^{(n)}|\ge d/2,\quad |z^*_{0,n}-h_j^{(n)}|\ge d/2.
\]
This and (\ref{razn_lam1}) yield (\ref{razn_lam}).

Consider the contour $C_R$ of the Fig.1 and
\begin{figure}

  \includegraphics[width=3.5 in, height=2.3 in]{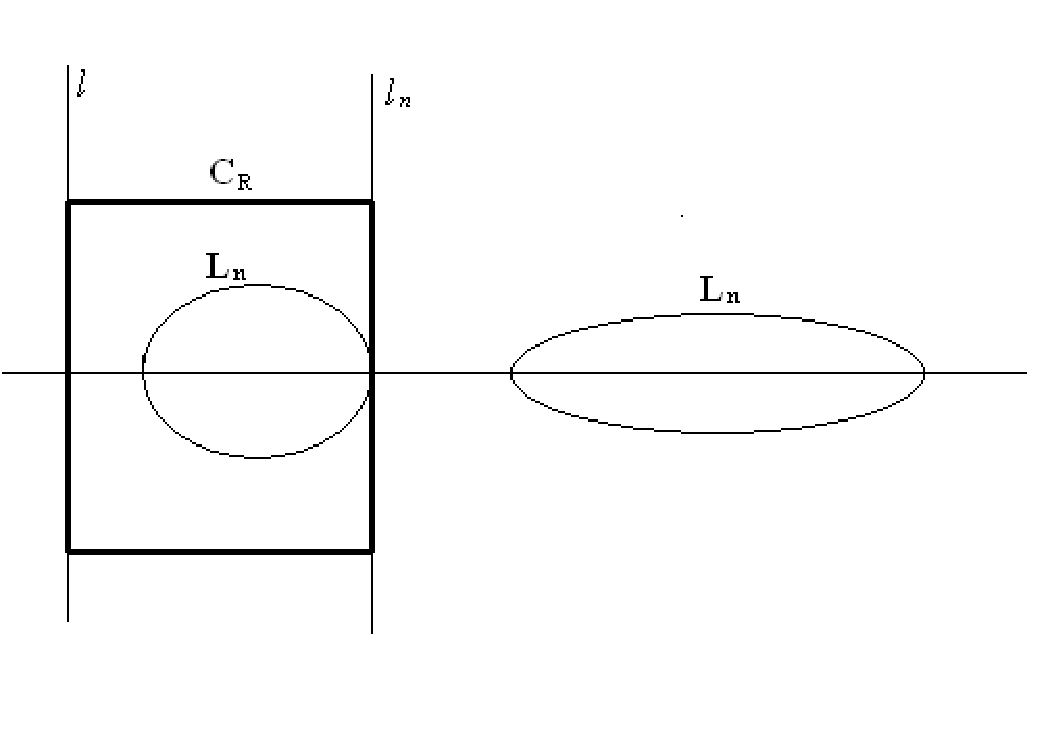}

\caption{Graph of the contour $C_R$.}
\label{fig:1}
\end{figure}

\begin{equation}\label{int_i_n}
\oint\limits_{L_n}\displaystyle\frac{dv}{2\pi}I_n(v),
\end{equation}
where
\begin{eqnarray}\label{I_n}
I_n(v)&=& -\displaystyle\oint\limits_{C_R}\displaystyle\frac{dt
}{2\pi}\exp\{n^{1/3}(v\xi-t\eta)+n(\lambda_{0,n}-\lambda_0)(t-v)\}\\ \notag
&&\times\displaystyle\frac{\exp\{n(S_n(t,\lambda_{0,n})- S_n(v,\lambda_{0,n}))\}} {v-t}
\end{eqnarray}
and the integral is understood in the Cauchy sense for $v=z^*_{0,n}$. We have
\[
I_n(v)=\left\{
\begin{array}{cl}
0&,v\,\,\hbox{is outside}\,\,C_R,\\
\dfrac{i}{2}\exp\{v(\eta-\xi)\}&,v=z^*_{0,n}
\end{array}
\right.
\]
(note that in view of Lemma \ref{l:min_max} $L_n$ and $l_n$ have only one point of intersection
$z^*_{0,n}$). We obtain
\begin{equation}  \label{Ker_osn}
\theta(\xi,\eta)\mathcal{K}_n(\xi,\eta)=\\
\displaystyle\int\limits_{l_n} \oint\limits_{L_n}
\mathcal{F}_n(t,v;\xi,\eta)d\,t\,d\,v,
\end{equation}
where
\begin{eqnarray}  \label{F_cal}
\mathcal{F}_n(t,v;\xi,\eta)&=&-\dfrac{n^{1/3}}{4\pi^2}
\exp\{n^{1/3}((v-z^*_{0,n})\xi-(t-z^*_{0,n})\eta)\}\\ \notag
&\times&e^{n(\lambda_{0,n}-\lambda_0)(t-v)}
\displaystyle\frac{\exp\{n(S_n(t,\lambda_{0,n})- S_n(v,\lambda_{0,n}))\}}{v-t },
\end{eqnarray}
\begin{equation}\label{theta1}
\theta(\xi,\eta)=\exp\{-n^{1/3}(\xi-\eta)z^*_{0,n}\},
\end{equation}
and $L_n$ and $l_n$ are defined in (\ref{L_n}) and (\ref{l_n}) respectively.

Now we need
\begin{lemma}\label{l:vne_okr}
There exists a sufficiently small $\delta>0$ such that for any $v\in L_n$ satisfying $|v-z^*_{0,n}|\ge\delta$ and any
$t\in l_n$ satisfying $|t-z^*_{0,n}|\ge\delta$ we have for $n>n_0$
\begin{equation*}
\Re S_n(v,\lambda_{0,n})\ge C_1,\quad \Re S_n(t,\lambda_{0,n})<-C_2,
\end{equation*}
where $C_1$ and $C_2$ do not depend on $\delta$.
\end{lemma}
The lemma is proved in Section 3.
Taking into account (\ref{eqv_f,pr}) and (\ref{S*}), we have for $t=z_{0,n}^*+iy\in l_n$
\begin{equation}\label{ots_besk}
\begin{array}{c}
\Re S_n(t,\lambda_{0,n})=-\dfrac{y^2}{2}+\dfrac{1}{n}\sum\limits_{j=1}^n\log \left|1+\dfrac{t-z_{0,n}^*}{z_{0,n}^*-
h_j^{(n)}}\right|\\
\le -\dfrac{y^2}{2}+\dfrac{1}{n}\sum\limits_{j=1}^n \left|\dfrac{t-z_{0,n}^*}{z_{0,n}^*-
h_j^{(n)}}\right| \le-\dfrac{y^2}{2}\\+|y|\left(\dfrac{1}{n}\sum\limits_{j=1}^n \dfrac{1}{(z_{0,n}^*-
h_j^{(n)})^2}\right)^{1/2}=-\dfrac{y^2}{2}+|y|\le -\dfrac{y^2}{4}
\end{array}
\end{equation}
for $|y|>C$, where $C$ is big enough.
Let us prove that for $|y|\ge\delta$
\begin{equation}\label{distLC}
\hbox{dist}\,(z_{0,n}^*+iy,L_n)\ge C\delta^2.
\end{equation}
Indeed, if $v\in L_n,\,\Im v\ge 0$ and $|v-z_{0,n}^*|\ge \delta$, then (\ref{Re}) (see below) yields
\begin{equation}\label{distCL}
\hbox{dist}\,(v, l_n)\ge \delta^2/3.
\end{equation}
Take $v\in L_n,\,\Im v\ge 0$, $|v-z_{0,n}^*|\le \delta$. We have from (\ref{Re}) (see below)
\[
\Im v(x)=s^{1/2}(x)\le C\sqrt{|x-z_{0,n}^*|}.
\]
Hence, $L_n$ lies below the curve $y=C\sqrt{|x-z_{0,n}^*|}$ and
\[
\hbox{dist}^2(z_{0,n}^*+iy,L_n)\ge \inf\limits_x\left\{\left(y-C\sqrt{|x-z_{0,n}^*|}\right)^2+
(x-z_{0,n}^*)^2\right\}\ge C_1\delta^4.
\]
This and (\ref{distCL}) give (\ref{distLC}).

According to Lemma \ref{l:z*_0,n} $|z_{0,n}^*-z_0|\le n^{-1/3-\varepsilon}$, thus $z_{0,n}^*$ is
uniformly bounded in $n$, and we can write
\begin{equation}\label{prexp_c}
\Re ((v-z_{0,n}^*)\xi-(t-z_{0,n}^*)\eta)\le C\Re v,
\end{equation}
when $\xi\in [-M,M]$.

Hence, (\ref{razn_lam}), Lemma \ref{l:vne_okr}, and (\ref{ots_besk}) -- (\ref{prexp_c}) imply
\begin{equation}\label{Int1}
\begin{array}{c}
\left|\left(\displaystyle\int\limits_{U_2}\int\limits_{L_n\backslash
U_1}+\displaystyle \int\limits_{l_n\backslash U_2} \displaystyle\int\limits_{L_n}\right)
\mathcal{F}_n(t,v;\xi,\eta)d\,t\,d\,v
\right|\le
n^{1/3}|L_n|\exp\{-Cn+cn^{1/3}\},
\end{array}
\end{equation}
where $\mathcal{F}_n(t,v;\xi,\eta)$ is defined in (\ref{F_cal}),
\begin{equation}\label{U_1,2}
U_1=\{z\in L_n:\, |z-z^*_{0,n}|\le\delta\}, \quad U_2=\{z\in l_n:\, |z-z^*_{0,n}|\le\delta\}
\end{equation}
and $|L_n|$ is the length of $L_n$.

Use now the assertion (see \cite[Lemma 6]{TSh:09}):
\begin{lemma}\label{l:dl_kont}
Let $l(x)$ be the oriented length of the upper part of the contour $L_n$  between $x_0=x_n(\lambda_0)$ and
$x$ (we take $l(x)>0$ for $x>x_0$ to obtain $l^\prime(x)>0$). Then for any collection
$\{h_j^{(n)}\}_{j=1}^n$, $l(x)$ admits the bound
$$|l(x_1)-l(x_2)|\le C|x_1-x_2|$$
with an absolute constant $C$. Moreover,
\begin{equation*}
|L_n|\le C n,
\end{equation*}
where $|L_n|$ is the length of $L_n$.
\end{lemma}
This lemma and (\ref{Int1}) yield
\begin{equation}  \label{Ker1}
\dfrac{\theta(\xi,\eta)}{n^{2/3}}K_n(\lambda_0+\xi/n^{2/3},\lambda_0+\eta/n^{2/3})\\
=\displaystyle\int\limits_{U_2}
\int\limits_{U_1}\mathcal{F}_n(t,v;\xi,\eta)d\,t\,d\,v+O(e^{-Cn}),
\end{equation}
where $\mathcal{F}_n$, $\theta(\xi,\eta)$ and
$U_1, U_2$ are defined in (\ref{F_cal}), (\ref{theta1}), and (\ref{U_1,2}) respectively.

This reduces (\ref{limK}) (and thus (\ref{Un})) to the relation
\begin{equation}\label{lim_I_ok}
\displaystyle\int\limits_{U_2}
\int\limits_{U_1}\mathcal{F}_n(t,v;\xi,\eta)d\,t\,d\,v=A(\gamma^{2/3} \xi,\gamma^{2/3}
\eta)+o(1),\,\,n\to\infty,
\end{equation}
where $\gamma$, $A$  are defined in (\ref{gam}) and (\ref{A}) respectively.

Taking into account (\ref{lam_0,n}) -- (\ref{S*}),  and (\ref{eqv_f,pr}), we get
\[
S_n(z^*_{0,n},\lambda_{0,n})=\dfrac{d}{d\,z}S_n(z^*_{0,n},\lambda_{0,n})=\dfrac{d^2}{d\,z^2}S_n(z^*_{0,n},\lambda_{0,n})=0,
\]
hence we obtain for $z\in L_n$ satisfying $|z-z^*_{0,n}|\le\delta$
\begin{equation}\label{S_okr}
S_n(z,\lambda_{0,n})=\dfrac{1}{n}\sum\limits_{j=1}^n\dfrac{1}{(z^*_{0,n}-h_j^{(n)})^3}\cdot\dfrac{(z-z^*_{0,n})^3}{3}
+O(\delta^4), \quad \delta\to 0.
\end{equation}
According to (\ref{pol_ga_n}) we have for $n>n_0$
\[
\dfrac{1}{n}\sum\limits_{j=1}^n\dfrac{1}{(z^*_{0,n}-h_j^{(n)})^3}>C>0.
\]
Thus, we can write for $z$ satisfying $|z-z^*_{0,n}|\le\delta$
\begin{equation}\label{s_sm}
S_n(z,\lambda_{0,n})=\gamma_n^{-2}\chi^3(z)/3,
\end{equation}
where $\chi(z)$ is analytic in the $\delta$-neighborhood of $z^*_{0,n}$ with the analytic
inverse $z(\varphi)$ (we choose $\chi(z)$ such that $\chi(z)\in\mathbb{R}$ for $z\in\mathbb{R}$) and
\begin{equation}\label{gamma_n}
\gamma_n=\left(\dfrac{1}{n}\sum\limits_{j=1}^n\dfrac{1}{(z^*_{0,n}-h_j^{(n)})^3}\right)^{-1/2}.
\end{equation}
Changing variables to $v=z(\varphi_1)$, $t=z(\varphi_2)$, rewrite the l.h.s. of (\ref{lim_I_ok}) as
\begin{equation}\label{I1_2}
\displaystyle\int\limits_{U_2}
\int\limits_{U_1}\mathcal{F}_n(t,v;\xi,\eta)d\,t\,d\,v=\displaystyle\int\limits_{U_2(\varphi)}
\int\limits_{U_1(\varphi)}\widetilde{\mathcal{F}}_n(\varphi_1,\varphi_2;\xi,\eta) d\varphi_2\,d\varphi_1,
\end{equation}
where
\begin{equation}\label{F_cal1}
\begin{array}{c}
\widetilde{\mathcal{F}}_n(\varphi_1,\varphi_2;\xi,\eta)=-\dfrac{n^{1/3}}{4\pi^2}
e^{n^{1/3}((z(\varphi_1)-z^*_{0,n})\xi-(z(\varphi_2)-z^*_{0,n})\eta)}\\
\times z^\prime(\varphi_1)z^\prime(\varphi_2)e^{n(\lambda_{0,n}-\lambda_0) (z(\varphi_2)-z(\varphi_1))}
 \displaystyle\frac{\exp\{n\gamma_n^{-2}(\varphi^3_2-\varphi^3_1)\}}
 {z(\varphi_1)-z(\varphi_2)},
\end{array}
\end{equation}
and
\begin{equation}\label{U_phi}
U_1(\varphi)=\{\varphi\in\mathbb{C}|z(\varphi)\in U_1\},\quad
U_2(\varphi)=\{\varphi\in\mathbb{C}|z(\varphi)\in U_2\}.
\end{equation}
Moreover, we have from (\ref{s_sm})
\begin{equation}\label{s_0}
\chi(z^*_{0,n})=0,\quad \dfrac{d}{d\,z}\chi(z_{0,n}^*)=1,
\end{equation}
hence
\begin{equation}\label{ogr_pr_phi}
0<C_1<|\chi^\prime(z)|<C_2, \quad |z-z_{0,n}^*|\le\delta.
\end{equation}
If $\sigma=\{z\in\mathbb{C}:|z-z^*_{0,n}|\le \delta\}$, then $\chi(\partial\sigma)$ is a closed
curve encircling $\varphi=0$ and lying between the circles
$\sigma_1=\{\varphi\in\mathbb{C}:|\varphi|=C_1\delta\}$ and
$\sigma_2=\{\varphi\in\mathbb{C}:|\varphi|=C_2\delta\}$ for $0<C_1<C_2$.
We have from (\ref{s_0})
\begin{equation}\label{z_pr}
\chi(0)=z_{0,n}^*,\quad \chi^\prime(0)=1, \quad 0<C_1<|\chi^\prime(\varphi)|<C_2,\quad \varphi\in \chi(\sigma).
\end{equation}
According to Lemma \ref{l:min_max}, $\Re S_n(z,\lambda_{0,n})\ge 0$ for $z\in U_1$ and we get $\Re \varphi^3_1\ge
0$ for $\varphi_1\in U_1(\varphi)$, i.e.,
\[
\cos (3\arg\varphi_1)\ge 0, \quad \varphi_1\in U_1(\varphi),
\]
where $U_1(\varphi)$ is defined in (\ref{U_phi}).
Hence, $U_1(\varphi)$ can be located only in sectors
\[
-\pi/6\le \arg\varphi\le\pi/6, \quad
\pi/2\le \arg\varphi\le 5\pi/6,\quad
7\pi/6\le \arg\varphi\le 3\pi/2.
\]
Besides, $\chi$ is conformal in $\sigma$ (see (\ref{ogr_pr_phi})), hence angle-preserving.
Taking into account that $\chi(z)\in\mathbb{R}$ for $z\in\mathbb{R}$, the angle between $L_n$
and the real axis at the point $z^*_{0,n}$ is $\pi/2$, and that $U_1(\varphi)$ is a continuous curve, we obtain that
$U_1(\varphi)$ can be located only in sectors
\begin{equation}\label{sectors}
\pi/2\le \arg\varphi\le 5\pi/6,\quad
7\pi/6\le \arg\varphi\le 3\pi/2.
\end{equation}
Note that we can take any curve $\widetilde{L}_1(\varphi)$ instead of $U_1(\varphi)$ provided that
$\widetilde{L}_1(\varphi)$ and $L_n\setminus U_1$ are "glued", i.e., the union of $z(L_1(\varphi))$ and
$L_n\setminus U_1$ form a closed contour encircling $\{h_j^{(n)}\}_{j=1}^n$.
Let us take
\begin{multline}
\widetilde{L}_1(\varphi)=\{\varphi\in\mathbb{C}:\arg \varphi=2\pi/3,\,\varphi\in \chi(\sigma)\}\\
\cup\{\varphi\in\mathbb{C}:\arg \varphi=4\pi/3,\,\varphi\in \chi(\sigma)\}
\cup L_{1,\delta}\cup L_{2,\delta},
\end{multline}
where $\sigma=\{z\in\mathbb{C}:|z-z_{0,n}^*|\le\delta\}$, $L_{1,\delta}$ is a curve along $\chi(\partial \sigma)$
from the point of intersection
of the ray $\arg\varphi=2\pi/3$ and $\chi(\partial \sigma)$ to the point $\varphi_{1,\delta}$ of intersection
of  $U_1(\varphi)$ and $\chi(\partial \sigma)$ ($\pi/2<\arg\varphi_{1,\delta}<5\pi/6$),
and $L_{2,\delta}$ is a curve along $\chi(\partial \sigma)$ from the point of intersection
of the ray $\arg\varphi=4\pi/3$ and $\chi(\partial \sigma)$ to the point $\varphi_{2,\delta}$ of intersection
of $U_1(\varphi)$ and $\chi(\partial \sigma)$ ($7\pi/6<\arg\varphi_{2,\delta}<3\pi/2$) (see Fig 2).

\begin{figure}

  \includegraphics[width=4 in, height=2.5 in]{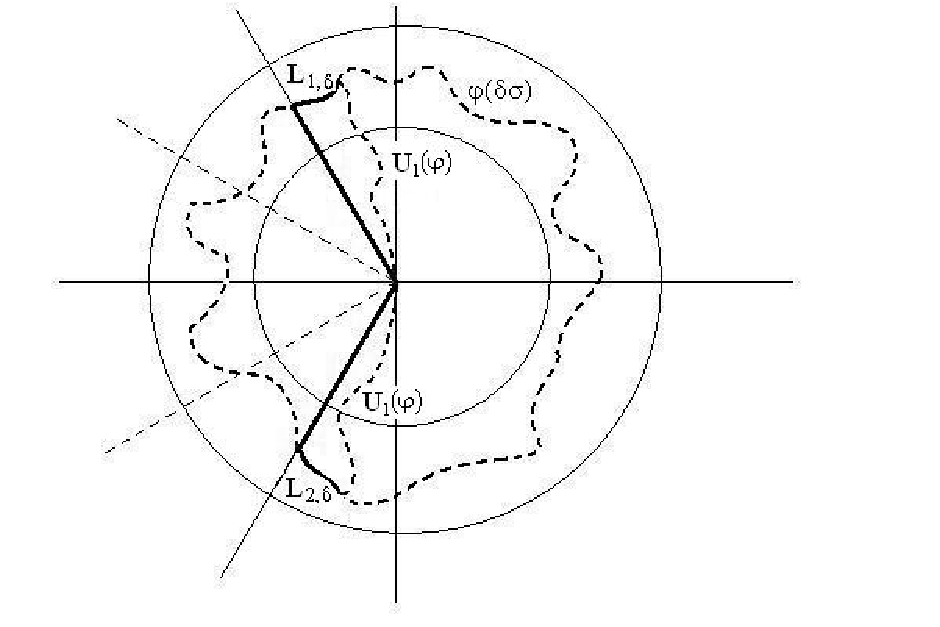}

\caption{Graph of $\widetilde{L}_1(\varphi)$.}
\label{fig:2}       
\end{figure}


  According to Lemma \ref{l:vne_okr} and (\ref{s_sm}), $\Re \varphi^3_{1,\delta}=r^3\cos 3\varphi_0>C>0$, where
$r=|\varphi_{1,\delta}|$, $\varphi_0=\arg\varphi_{1,\delta}$. Since $0<C_1<r<C_2$, we have
\[
\cos 3\varphi_0\ge C/C_2^3>0.
\]
Moreover, it is easy to see that $\cos (3\arg\varphi_1)>\cos 3\varphi_0$ along $L_{1,\delta}$ (since
$\cos 3x$ is monotone increasing for $x\in[\pi/2,2\pi/3]$ and monotone decreasing for
$x\in [2\pi/3,5\pi/6]$).
This and $|\varphi_1|>C_1$ imply for $\varphi_1\in L_{1,\delta}$
\[
\Re \left(\dfrac{\gamma_n^{-2}\varphi_1^3}{3}\right)>C>0,\quad \varphi_1\in L_{1,\delta}.
\]
Also we have from (\ref{z_pr})
\[
|z(\varphi_1)-z^*_{0,n}|\le C_2|\varphi_1|<C,\quad\varphi_1\in \chi(\sigma).
\]
This, (\ref{razn_lam}) and  (\ref{z_pr}) yield
\begin{equation}\label{ots_L_d}
\left|\displaystyle\int\limits_{U_2(\varphi)}\int\limits_{L_{1,\delta}}
\widetilde{\mathcal{F}}_n(\varphi_1,\varphi_2;\xi,\eta)\,d\varphi_1\,d\varphi_2\right|
 \le  Cn^{1/3}\exp\{-Cn+cn^{1/3}\},
\end{equation}
where $\widetilde{\mathcal{F}}_n(\varphi_1,\varphi_2;\xi,\eta)$ is defined in (\ref{F_cal1}).
Similarly, we can prove that integral over
$L_{2,\delta}$ does not contribute to the l.h.s. of (\ref{lim_I_ok}).

We have shown that integral over $U_1(\varphi)$ in (\ref{I1_2}) can be replaced
to the integral over the contour
\begin{equation}\label{l^{(1)}}
l^{(1)}=\{\varphi\in\mathbb{C}:\arg \varphi=2\pi/3,\,\varphi\in \chi(\sigma)\}
\cup\{\varphi\in\mathbb{C}:\arg \varphi=4\pi/3,\,\varphi\in \chi(\sigma)\},
\end{equation}
i.e.
\begin{equation}\label{vsp1}
\displaystyle\int\limits_{U_2}\int\limits_{U_1}\mathcal{F}_n(t,v;\xi,\eta)d\,t\,d\,v=
\displaystyle\int\limits_{U_2(\varphi)}\int\limits_{l^{(1)}}
\widetilde{\mathcal{F}}_n(\varphi_1,\varphi_2;\xi,\eta)\,d\varphi_1\,d\varphi_2+O(e^{-Cn}).
\end{equation}
The same argument implies that the integral over $U_2(\varphi)$ on the l.h.s. of (\ref{lim_I_ok}) can be
replaced by the integral over the contour
\begin{equation}\label{l^{(2)}}
l^{(2)}=\{\varphi\in\mathbb{C}:\arg \varphi=\pi/3,\,\varphi\in \chi(\sigma)\}
\cup\{\varphi\in\mathbb{C}:\arg \varphi=5\pi/3,\,\varphi\in \chi(\sigma)\}.
\end{equation}
Indeed, we use Lemma \ref{l:min_max} to obtain $\Re \varphi^3_2\le 0$ for $\varphi_2\in U_2(\varphi)$ and thus
$U_2(\varphi)$ can be located only in sectors
\[
\pi/6\le \arg\varphi\le\pi/2, \quad
5\pi/6\le \arg\varphi\le 7\pi/6,\quad
3\pi/2\le \arg\varphi\le 11\pi/6.
\]
Using again that $\chi(z)$ is conformal in $\sigma$, we obtain that
$U_2(\varphi)$ can be located only in sectors
\begin{equation*}
\pi/6\le \arg\varphi\le \pi/2,\quad
3\pi/2\le \arg\varphi\le 11\pi/6.
\end{equation*}
Now we can replace the integral over $U_2(\varphi)$ by the integral over
\begin{multline*}
\widetilde{L}_2(\varphi)=\{\varphi\in\mathbb{C}:\arg \varphi=\pi/3,\,\varphi\in \chi(\sigma)\}\\
\cup\{\varphi\in\mathbb{C}:\arg \varphi=5\pi/3,\,\varphi\in \chi(\sigma)\}
\cup L_{3,\delta}\cup L_{4,\delta},
\end{multline*}
where $L_{3,\delta}$ is a curve along $\chi(\partial \sigma)$ from the point of intersection
of the ray $\arg\varphi=\pi/3$ and $\chi(\partial \sigma)$ to the point $\varphi_{3,\delta}$ of intersection
of  $U_2(\varphi)$ and $\chi(\partial \sigma)$ ($\pi/6<\arg\varphi_{3,\delta}<\pi/2$),
and $L_{4,\delta}$ is a curve along $\chi(\partial \sigma)$ from the point of intersection
of the ray $\arg\varphi=5\pi/3$ and $\chi(\partial \sigma)$ to the point $\varphi_{4,\delta}$ of intersection
of $U_2(\varphi)$ and $\chi(\partial \sigma)$ ($3\pi/2<\arg\varphi_{4,\delta}<11\pi/6$).
It follows from Lemma \ref{l:vne_okr} that we can replace $\widetilde{L}_2(\varphi)$ by the contour
$l^{(2)}$ of (\ref{l^{(2)}}).

Thus, (\ref{Ker1}), (\ref{I1_2}) and (\ref{vsp1}) imply
\begin{multline}  \label{Ker2}
\dfrac{\theta(\xi,\eta)}{n^{2/3}}K_n(\lambda_0+\xi/n^{2/3},\lambda_0+\eta/n^{2/3})\\
=\displaystyle\int\limits_{l^{(2)}}
\int\limits_{l^{(1)}}\widetilde{\mathcal{F}}_n(\varphi_1,\varphi_2;\xi,\eta)d\,\varphi_1\,d\,\varphi_2+O(e^{-Cn}),
\end{multline}
where $\widetilde{\mathcal{F}}_n(\varphi_1,\varphi_2;\xi,\eta)$ is defined in (\ref{F_cal1}), and to prove
(\ref{limK}) it suffices to show that
\begin{equation}\label{lim_I_ok1}
\displaystyle\int\limits_{l^{(2)}}
\int\limits_{l^{(1)}}\widetilde{\mathcal{F}}_n(\varphi_1,\varphi_2;\xi,\eta)d\,\varphi_1\,d\,\varphi_2=A(\gamma^{2/3}\xi,\gamma^{2/3}\eta)+o(1),
\end{equation}
where $l^{(1)}$ and $l^{(2)}$ are defined in (\ref{l^{(1)}}), (\ref{l^{(2)}}).

According to the choice of $l^{(1)}$ and $l^{(2)}$, we have
\begin{equation}\label{phi^3}
\begin{array}{c}
\Re \varphi^3=r^3,\quad \varphi\in l^{(1)},\\
\Re \varphi^3=-r^3,\quad \varphi\in l^{(2)},
\end{array}
\end{equation}
where $r=|\varphi|$.

Now set $$\sigma_n=\{\varphi\in\mathbb{C}:|\varphi|\le \log n/n^{1/3}\}.$$ It is easy to see that
 $\sigma_n\subset \chi(\sigma)$.  Taking into account (\ref{razn_lam}), (\ref{z_pr}), and (\ref{phi^3}),
 we obtain for $\varphi_1\in
l^{(1)}\setminus \sigma_n$, $\varphi_2\in l^{(2)}$
\begin{equation}\label{ots_l_n}
\left|\widetilde{\mathcal{F}}_n(\varphi_1,\varphi_2;\xi,\eta)\right|
 \le Cn^{1/3}\exp\{-nC_1r^3+n^{1/3}C_2r\},
\end{equation}
where $r=|\varphi_1|\ge \dfrac{\log n}{n^{1/3}}$. Since $n^{1/3}r\ge \log n$ for $\varphi_1\in
l^{(1)}\setminus \sigma_n$, the integral over $l^{(1)}\setminus \sigma_n$ is $O(e^{-Cn})$ as $n\to \infty$.
Similarly,  the integral over $l^{(2)}\setminus \sigma_n$ is $O(e^{-Cn})$ as $n\to \infty$.
It suffices to prove that
\begin{equation}\label{lim_I_ok2}
\begin{array}{c}
I:=\displaystyle\int\limits_{l_{2,n}}
\int\limits_{l_{1,n}}\widetilde{\mathcal{F}}_n(\varphi_1,\varphi_2;\xi,\eta)d\,\varphi_1\,d\,\varphi_2=A(\gamma^{2/3}\xi,\gamma^{2/3}\eta)+o(1),
\end{array}
\end{equation}
where $l_{1,n}=l^{(1)}\cap\sigma_n$, $l_{2,n}=l^{(2)}\cap\sigma_n$.

We have from (\ref{z_pr}) for $\varphi\in\sigma_n$
\begin{equation*}
\begin{array}{c}
z(\varphi)=z^*_{0,n}+\varphi+O(\log^2n/n^{2/3}),\,\,n\to\infty,\\
z^\prime(\varphi)=1+O(\log n/n^{1/3}),\,\,n\to\infty.
\end{array}
\end{equation*}
Hence, (\ref{razn_lam}) implies
\begin{multline}\label{pered_zam}
\widetilde{\mathcal{F}}_n(\varphi_1,\varphi_2;\xi,\eta)=
\exp\{n^{1/3}(\varphi_1\xi-\varphi_2\eta)\}\\
\times \dfrac{\exp\{n\gamma_n^{-2}(\varphi^3_2-\varphi^3_1)\}}
 {\varphi_1-\varphi_2}(1+o(1)),\,\,n\to\infty.
\end{multline}
Changing variables in (\ref{lim_I_ok2}) as
$\gamma_n^{-2/3} n^{1/3}\varphi_1\to i\varphi_1$, $\gamma_n^{-2/3} n^{1/3}\varphi_2\to i\varphi_2$,
we obtain
\begin{equation*}
I=\displaystyle\int\limits_{\widetilde{l}_{2,n}}
\int\limits_{\widetilde{l}_{1,n}}
F(\varphi_1,\varphi_2;\xi,\eta)(1+o(1))\,d\varphi_1\,d\varphi_2,
\end{equation*}
where
\begin{equation}\label{F}
F(\varphi_1,\varphi_2;\xi,\eta)=\dfrac{\gamma_n^{2/3}}{4\pi^2}\exp\{i\gamma_n^{2/3}(\varphi_1\xi-\varphi_2\eta)\}
 \displaystyle\frac{\exp\{-i\varphi^3_2/3+i\varphi^3_1/3\}}
 {i\varphi_2-i\varphi_1}
\end{equation}
and
$$
\begin{array}{c}
\widetilde{l}_{1,n}=\{\varphi\in \mathbb{C}:\arg\varphi=\pi/6\,\,\hbox{or}\,\,5\pi/6 ,\,|\varphi|\le \gamma_n^{-2/3}\log
n\},\\
\widetilde{l}_{2,n}=\{\varphi\in \mathbb{C}:\arg\varphi=-\pi/6\,\,\hbox{or}\,-5\pi/6,\,|\varphi|\le \gamma_n^{-2/3}\log n\}.
\end{array}
$$
Note that if $\varphi_1$ and $\varphi_2$ satisfy $\arg \varphi_1=\pi/6\,\,\hbox{or}\,\,5\pi/6$,
$|\varphi_1|>\gamma_n^{-2/3}\log n$
and $\arg\varphi_2=-\pi/6\,\,\hbox{or}\,-5\pi/6$, then we have
\[
|\varphi_1-\varphi_2|> \frac{\sqrt{3}\log n}{2\gamma_n^{2/3}},\quad
\Re(i\varphi^3_1/3+i\gamma_n^{2/3}\varphi_1\xi)\le -\gamma_n^{-2}\log^3 n/3,
\]
and we get in view of the inequality $0<C_1<\gamma_n<C_2$
\begin{equation*}
\begin{array}{c}
\left|\displaystyle\int\limits_{\widetilde{l}_{2}}
\int\limits_{\widetilde{l}_1\setminus\widetilde{l}_{1,n}}
F(\varphi_1,\varphi_2;\xi,\eta)(1+o(1))\,d\varphi_1\,d\varphi_2\right|\\
 \le \dfrac{Ce^{-\gamma_n^{-2}\log^3 n/6}}{\log n}\displaystyle\int\limits_{\widetilde{l}_{2}}
\int\limits_{\widetilde{l}_1\setminus\widetilde{l}_{1,n}}
e^{i\gamma_n^{2/3}(\varphi_1\xi-\varphi_2\eta)+(i\varphi^3_1-i\varphi^3_2)/3}d\varphi_1d\varphi_2\\
\le C e^{-\gamma_n^{-2}\log^3 n/6},
 \end{array}
\end{equation*}
where
$$
\begin{array}{lll}
\widetilde{l}_{1}&=&\{\varphi\in \mathbb{C}:\arg\varphi=\pi/6\,\,\hbox{or}\,\,5\pi/6\},\\
\widetilde{l}_{2}&=&\{\varphi\in \mathbb{C}:\arg\varphi=-\pi/6\,\,\hbox{or}\,-5\pi/6\}.\\
\end{array}
$$
The same bound holds for the integral over $\widetilde{l}_2\setminus \widetilde{l}_{2,n}$.

We have as $n\to\infty$
\begin{equation}  \label{Ker3}
\theta(\xi,\eta)\mathcal{K}_n(\xi,\eta)\\
=\displaystyle\int\limits_{\widetilde{l}_{1}}
\int\limits_{\widetilde{l}_{2}}F(\varphi_1,\varphi_2;\xi,\eta)(1+o(1))d\,\varphi_1\,d\,\varphi_2+O(e^{-C\log^3n}),
\end{equation}
where $\mathcal{K}_n(\xi,\eta)$ is defined in (\ref{K_kal}).
To prove (\ref{Un}) it remains to show that
\begin{equation}\label{int_okr}
\displaystyle\int\limits_{\widetilde{l}_{1}}
\int\limits_{\widetilde{l}_{2}}F(\varphi_1,\varphi_2;\xi,\eta)(1+o(1))d\,\varphi_1\,d\,\varphi_2
 =A(\gamma^{2/3}\xi,\gamma^{2/3}\eta)+o(1).
\end{equation}
Writing
\[
e^{-i\varphi_2a+i\varphi_1b}=\dfrac{i}{a-b}\left(\dfrac{\partial}{\partial\varphi_1}+
\dfrac{\partial}{\partial\varphi_2}\right)e^{-i\varphi_2a+i\varphi_1b},
\]
plugging this and (\ref{F}) in the l.h.s. of (\ref{int_okr}) and integrating by parts, we obtain in view of (\ref{A}) -- (\ref{Ai})
\begin{multline}\label{end}
\displaystyle\int\limits_{\widetilde{l}_{2}}\int\limits_{\widetilde{l}_{1}}
F(\varphi_1,\varphi_2;\xi,\eta)d\,\varphi_1\, d\,\varphi_2
\\=\displaystyle\int\limits_{\widetilde{l}_{2}}\int\limits_{\widetilde{l}_{1}}
\dfrac{\varphi_1+\varphi_2}{\eta-\xi}e^{i\gamma_n^{2/3}(\varphi_1\xi-\varphi_2\eta)+i(\varphi^3_1-\varphi^3_2)/3}
\dfrac{i\,d\varphi_1\, d\varphi_2}{4\pi^2}\\
=\dfrac{\hbox{Ai}(\gamma_n^{2/3}\xi)\hbox{Ai}^\prime(\gamma_n^{2/3}\eta)-\hbox{Ai}^\prime(\gamma_n^{2/3}\xi)\hbox{Ai}(\gamma_n^{2/3}\eta)}{\gamma_n^{2/3}
(\xi-\eta)}\\=A(\gamma_n^{2/3}\xi,\gamma_n^{2/3}\eta).
\end{multline}
Conditions (i) -- (ii) of Theorem \ref{thm:1} and Lemma \ref{l:z*_0,n} yield
\begin{equation}\label{lim_gam}
\lim\limits_{n\to\infty}\gamma_n=\gamma,
\end{equation}
where $\gamma$ and $\gamma_n$ are defined in (\ref{gam}), (\ref{gamma_n}) respectively.
Hence, (\ref{Un}) is proved.

\textit{Remarks}

1. All the bounds in the proofs of results of this section hold if we take $|\xi|,|\eta|\le c n^{2/3}$
for a sufficiently small $c>0$.

2. Formulas (\ref{Ker3}) for $\xi=\eta=-c n^{2/3}$ with a sufficiently small $c>0$, (\ref{end}), (\ref{lim_gam}) and the
asymptotic formula (see \cite{Ab-St:65})
\[
A(x,x)=\dfrac{1}{\pi}\sqrt{-x}(1+o(1)),\quad x\to-\infty
\]
implies (\ref{gamma}).

It is well-known (see e.g. \cite{Me:91}) that
\begin{multline}\label{gap}
E_{n}\left(\Delta_n\right)
=1
+\sum\limits_{l=1}^\infty\dfrac{(-1)^l}{l!}\\ \times
\displaystyle\int\limits_{\triangle^l}\det\left\{(\gamma)^{-2/3}\mathcal{K}_n\left(x_i/\gamma^{2/3},
x_j/\gamma^{2/3}\right)\right\}_{i,j=1}^l\prod\limits_{j=1}^ld\,x_j,
\end{multline}
where $\Delta=[a,b]$ and $\Delta_n=[\lambda_0+a/(\gamma n)^{2/3},\lambda_0+b/(\gamma n)^{2/3}]$.
Since $A(\xi,\eta)$ is uniformly bounded in $\xi,\eta\in [-M,M]$,
according to the dominant convergence theorem, (\ref{Un}) yields (\ref{gp}) for $a,b\in [-M,M]$.

To prove (\ref{gp}) for $b=+\infty$ we need an additional bound on the $K_n(\lambda,\lambda)$
\begin{lemma}\label{l:as_ker}
There exists $n_0$ such that we have for $n>n_0$
\[
\left|K_n(\lambda,\lambda)\right|\le e^{-Cn},\quad \lambda\in
\mathbb{R}\setminus\mathrm{supp}\,N.
\]
Moreover, if $\lambda$ is big enough, then
\begin{equation}\label{big_lam}
\left|K_n(\lambda,\lambda)\right|\le e^{-n\lambda^2/4},\quad n>n_0.
\end{equation}
\end{lemma}
The lemma is proved in the next Section.
The lemma, the asymptotic formula
\[
A(\xi,\eta)=C_1e^{-C_2\xi^{3/2}}(1+o(1)),\,\xi\to +\infty.
\]
following from those for the Airy function, and (\ref{Ker3}) imply
\begin{equation}  \label{Ker4}
\theta(\xi,\eta)\mathcal{K}_n(\xi,\eta)\\
=C_1e^{-C_2\xi^{3/2}}(1+o_n(1))(1+o_\xi(1))+O(e^{-C\log^3n}).
\end{equation}
This and (\ref{gap}) yield (\ref{gp}) for $b\le n^{2/3}\delta$ with a sufficiently small $\delta$.

Take now $\Delta=[\lambda_0+\xi/n^{2/3}, b]$, $\Delta_1=[\lambda_0+\xi/n^{2/3}, \lambda_0+\delta]$,
 where $|\xi|\le M$ and $\delta$ is small enough. Set
\begin{eqnarray}\notag
P_1&=&\mathbf{P}\{\lambda_j^{(n)}\not\in \Delta,\,j=1,..,n\},\quad
P_2=\mathbf{P}\{\lambda_j^{(n)}\not\in \Delta_1,\,j=1,..,n\},\\ \notag
P_3&=&\mathbf{P}\{\exists j\in\{1,..,n\}:\lambda_j^{(n)}\in \Delta\setminus\Delta_1\}.
\end{eqnarray}
Then we have
\begin{equation}\label{ner_pr}
P_2-P_3\le P_1\le P_2.
\end{equation}
Since we prove (\ref{gp}) for $P_2$, we are left to prove that
\[
P_3\le e^{-Cn},\quad n\to\infty.
\]
This can be obtained from Lemma \ref{l:as_ker} by the inequality
\[
P_3\le n\mathbf{P}\{\lambda_1^{(n)}\in
\Delta\setminus\Delta_1\}=\int\limits_{\Delta\setminus\Delta_1}K_n(\lambda,\lambda)d\,\lambda\le
C_1ne^{-C_2n}<e^{-Cn}.
\]
Using the same arguments and (\ref{big_lam}) we obtain (\ref{gp}) for $b=+\infty$.

\section{Proof of auxiliary statements for \protect Theorem~\ref{thm:1}}
\textbf{Proof of Proposition \ref{p:lim_cont}.}

It was proved in \cite[Lemma 1]{TSh:09} that the limit $f(\lambda+i0)$ exists for all $\lambda\in\mathbb{R}$, the
equation (\ref{eqv_f_0}) is uniquely soluble, the limiting NCM $N$ is absolutely continuous, its density $\rho$ is continuous, and
$\Im f(\lambda+i0)=\pi\rho(\lambda)$. Since $\rho(\lambda_0)=0$ by the conditions of Theorem \ref{thm:1} we obtain
$z_0\in \mathbb{R}$. Thus, we are left to prove that $z_0$ is a solution of equation (\ref{eqv_f_0}) for
$\lambda=\lambda_0$ and that condition (\ref{eqv_z_0}) holds. The first assertion follows from
(\ref{Pe0}) and the condition (ii) of Theorem \ref{thm:1}. Since $\lambda_0$ is an edge of the spectrum,
the implicit function theorem yields that the derivative of (\ref{Pe0}) with respect to $f$ is
zero, which gives the first equality of (\ref{eqv_z_0}). Thus, we have for $V(z)$ of (\ref{eqv_f_0})
\[
V(z_0)-\lambda_0=\dfrac{d}{dz}V(z_0)=0.
\]
Set
\begin{equation}\label{z_lam}
z(\lambda)=\lambda+f(\lambda+i0).
\end{equation}
It follows from the result of \cite{TSh:09} and from (\ref{r_ep}) that
\[
z(\lambda)\in \mathbb{R},\quad z^\prime(\lambda)\ge 0, \quad \lambda\in (\lambda_0,\lambda_0+\delta],
\]
and that
\[
\dfrac{d}{dz}V(z(\lambda))\ge 0, \quad \lambda\in (\lambda_0,\lambda_0+\delta].
\]
We have for a sufficiently small $\delta_1>0$
\begin{equation}\label{pol}
\dfrac{d}{dz}V(x)\ge 0, \quad x\in(z_0,z_0+\delta_1].
\end{equation}
Hence, $\dfrac{d^2}{dz^2}V(z_0)\ge 0$. Besides, $\dfrac{d^3}{dz^3}V(z_0)<0$.
This yields that if $\dfrac{d^2}{dz^2}V(z_0)=0$, then $z_0$ is a maximum point of $\dfrac{d}{dz}V(z)$, $z\in
\mathbb{R}$, which contradicts with (\ref{pol}).
$\quad\Box$

\medskip

\textbf{Proof of Lemma \ref{l:z*_0,n}.}

Set $\omega_n=\{z: |z-z_0|\le n^{-1/3-\varepsilon}\}$ and $\omega=\{z:|z-z_0|\le\delta\}$,
where $0<\varepsilon<\alpha/2$,
$\alpha$ and $z_0$ are defined in (\ref{alpha}) and (\ref{z_0}), and $\delta$ is small enough. Consider the functions $\phi(z)=1-\dfrac{d}{d\,z}f^{(0)}(z)$ and
$\phi_n(z)=\dfrac{d}{d\,z}f^{(0)}(z)-\dfrac{d}{d\,z}f^{(0)}_n(z)$. Taking into account (\ref{eqv_z_0}) we have
\begin{eqnarray*}
\dfrac{d}{d\,z}f^{(0)}(z)&=&\dfrac{d}{d\,z}f^{(0)}(z_0)
+\dfrac{d^2}{d\,z^2}f^{(0)}(z_0)(z-z_0)
+O(n^{-2/3-2\varepsilon})\\
&=&1+\dfrac{d^2}{d\,z^2}f^{(0)}(z_0)(z-z_0)
+O(n^{-2/3-2\varepsilon}), \quad z\in\partial\omega_n,\,\,n\to\infty.
\end{eqnarray*}
Besides, we have from (\ref{eqv_z_0}) (recall that $z_0\in \mathbb{R}$)
\[
\dfrac{d^2}{d\,z^2}f^{(0)}(z_0)=\int\dfrac{2N^{(0)}(d\,h)}{(h-z_0)^3}<-C<0,
\]
hence
\begin{equation}\label{ogr_phi}
|\phi(z)|\ge C n^{-1/3-\varepsilon},\quad z\in\partial\omega_n,\quad n>n_0.
\end{equation}
In addition, it follows from the conditions (ii) -- (iii) of Theorem \ref{thm:1} that $h_j^{(n)}\not\in
\omega$, $j=1,..,n$ for $n>n_0$, thus $f^{(0)}$ and
$f^{(0)}_n$ are analytic in $\omega$, and condition (i) of Theorem \ref{thm:1} yields
\[
\left|\dfrac{d}{d\,z}f^{(0)}(z)-\dfrac{d}{d\,z}f^{(0)}_n(z)\right|\le n^{-1/3-(\alpha-\varepsilon)},\quad
z\in\partial\omega_n.
\]
This, (\ref{ogr_phi}), and the inequality $\varepsilon< \alpha/2$ yield for $n>n_0$
\[
|\phi(z)|>|\phi_n(z)|,\quad z\in\partial\omega_n.
\]
Both functions $\phi$ and $\phi_n$ are analytic in $\omega_n$, since we noted above that $f^{(0)}$ and $f^{(0)}_n$
are analytic in $\omega$. Hence, the Rouchet theorem implies that
$\phi(z)$ and $\phi(z)+\phi_n(z)=1- \dfrac{d}{d\,z} f^{(0)}_n(z)$ have the same number of zeros in
$\omega_n$. Since $\phi(z)$ has only one zero $z_0$ in $\omega_n$ (see Proposition \ref{p:lim_cont}),
we conclude that for any $n>n_0$ equation (\ref{eqv_f,pr}) has the unique solution $z^*_{0,n}$ in~$\omega_n$.
Moreover, since in view of condition (i) of Theorem \ref{thm:1} $f^{(0)}_n(z_{0,n}^*)\to f^{(0)}(z_0)$ as $n\to\infty$, we obtain (\ref{pol_ga_n}) from (\ref{eqv_z_0}).
Note that taking $\omega$ instead of $\omega_n$,
we can obtain analogously that equation (\ref{eqv_f,pr}) has only one solution in $\omega$.

 Similarly we can prove (\ref{z-z_0}). Indeed, consider two functions
 $$\psi(z)=-f^{(0)}(z)+z-\lambda_0,\quad \psi_n(z)=-f^{(0)}_n(z)+f^{(0)}(z),$$
where $f^{(0)}_n,\,f^{(0)}$ are defined in (\ref{f_0,n}), (\ref{eqv_f_0}). Since $z_0$ is a zero of
the multiplicity two of $\psi(z)$ (see (\ref{eqv_z_0})), we obtain
\begin{equation}
|\psi(z)|\ge C_0n^{-2/3-2\varepsilon},\quad z\in\partial\omega_n,
\end{equation}
where $C_0$ is a $n$-independent constant. Besides, we have for $n>n_0$ from the condition (i) of Theorem \ref{thm:1}
\[
|f^{(0)}(z)-f^{(0)}_n(z)|\le n^{-2/3-\alpha},\quad z\in \omega_n.
\]
Since $\varepsilon<\alpha/2$, we obtain for $n>n_0$
\begin{equation*}
|\psi(z)|>|\psi_n(z)|, \,\,z\in\partial\omega_n.
\end{equation*}
Both functions $\psi$ and $\psi_n$
are analytic in $\omega_n$, since we noted above that $f^{(0)}$ and $f^{(0)}_n$
are analytic in $\omega$. Hence, the Rouchet theorem implies that
$\psi(z)$ and $\psi(z)+\psi_n(z)=z- f^{(0)}_n(z)-\lambda_0$ have the same number of zeros in $\omega_n$.
Since $\psi(z)$ has only one zero of the multiplicity two in $\omega_n$, we conclude that for any $n>n_0$
equation (\ref{eqv_f_0,n}) with $\lambda=\lambda_0$ has two zeros in $\omega_n$. If one of these two zeros is
not real, then it is $z_n(\lambda_0)$ or $\overline{z_n(\lambda_0)}$ (since (\ref{eqv_f_0,n})
does not have any other zeros in $\mathbb{C}\setminus\mathbb{R}$) and hence (\ref{z-z_0}) is proved.
If both zeros are real, then since in view of (iii) of Theorem \ref{thm:1} $h_j^{(n)}\not\in\omega_n$, $j=1,..,n$,
there are no $h_j^{(n)}$-s between these zeros. If they lie to the left (right) of all $h_j^{(n)}$-s, then one of them
is $z_n(\lambda_0)$, since (\ref{eqv_f_0,n}) with $\lambda=\lambda_0$ has only two zeros there.
If they lie on a segment between adjacent $h_j^{(n)}$-s, then the segment contains three zeros $x_1<x_2<x_3$
and one of them is $z_n(\lambda_0)$. Since $\Re z_n^\prime(\lambda_0)>0$
(see Lemma \ref{l:min_max} below), we have $z_n(\lambda_0)=x_2$.  Thus, in this case $z_n(\lambda_0)$ also belongs
to $\omega_n$ (since if $x_1,x_3\in\omega_n$, then $x_2\in\omega_n$ too).
$\quad\Box$

\medskip

\textbf{Proof of Lemma \ref{l:vne_okr}.}

Let $x_n(\lambda)$ and $y_n(\lambda)$ be the real and imaginary parts of $z_n(\lambda)$. It follows from Lemma \ref{l:min_max}
that one can express $y_n(\lambda)$ via $x_n(\lambda)$ to obtain the "graph" $y_n(x)$ of the upper part of $L_n$. Denote
\begin{equation}\label{obozn}
\begin{array}{c}
y_n^2(x)=s(x),\quad x-h_j^{(n)}=\triangle_j,\,\,\\
\sigma_k= \dfrac{1}{n}\sumd\limits_{j=1}^n
 \dfrac{1}{(\triangle_j^2+s)^k},\quad
\sigma_{kl}= \dfrac{1}{n}\sumd\limits_{j=1}^n
\dfrac{\triangle^l_j}{(\triangle_j^2+s)^k},\,\, k=\overline{1,3},\,l=1,2,\\
\sigma_{k}^{(0)}= \dfrac{1}{n}\sumd\limits_{j=1}^n
\dfrac{1}{(z_{0,n}^*-h_j^{(n)})^k},\,\, k=\overline{1,4},
\end{array}
\end{equation}
and put $z\in L_n$, $x=\Re z$. Then we have from (\ref{S_n})
\begin{equation}\label{ReS_x}
\Re S_n(z,\lambda_{0,n})=\dfrac{x^2-s(x)}{2}+\dfrac{1}{2n}\sum\limits_{j=1}^n\log
((x-h_j^{(n)})^2+s(x))-\lambda_{0,n}x-S^*.
\end{equation}
Besides, taking the imaginary and real parts of (\ref{eqv_f_0,n}) we obtain for $x=\Re z$, $z\in L_n$
\begin{equation}\label{cond}
\dfrac{1}{n}\sumd\limits_{j=1}^n
 \dfrac{1}{\triangle_j^2+s}=1,\quad
x+\dfrac{1}{n}\sumd\limits_{j=1}^n
 \dfrac{\triangle_j}{\triangle_j^2+s}=\lambda.
\end{equation}
Differentiating the first equation in (\ref{cond}) with respect to $x$, we obtain the
equality
\begin{equation}\label{s_pr}
-s^\prime(x)\dfrac{1}{n}\displaystyle\sum\limits_{j=1}^n\dfrac{1}
{(\triangle_j^2+s(x))^2}-\dfrac{2}{n}\sumd\limits_{j=1}^n\dfrac{\triangle_j}
{(\triangle_j^2+s(x))^2}=0
\end{equation}
implying that for $z\in L_n$, $x=\Re z$
\begin{equation}\label{s1}
|s^\prime(x)|=2|\sigma_{21}|\sigma_2^{-1}\le
2\sigma_{22}^{1/2} \sigma_2^{-1/2}\le 2\sigma_2^{-1/2}\le 2\sigma_1^{-1}=2.
\end{equation}
Substituting $x=z^*_{0,n}$ in (\ref{s_pr}) we get
\begin{equation}\label{s_pr_0}
s^\prime(z^*_{0,n})=-\dfrac{2\sigma_{3}^{(0)}}{\sigma_{4}^{(0)}}.
\end{equation}
This, (\ref{S*}) and (\ref{ReS_x}) -- (\ref{cond}) imply
\begin{equation}\label{pr_ReS}
\begin{array}{c}
\Re S_n(z^*_{0,n},\lambda_{0,n})=0,\\
\dfrac{d}{d\,x}S_n(z^*_{0,n},\lambda_{0,n})=x+\sigma_{11}-\lambda_{0,n}\Big|_{x=z^*_{0,n}}=0,\\
\dfrac{d^2}{d\,x^2}S_n(z^*_{0,n},\lambda_{0,n})=2(\sigma_{3}^{(0)})^2(\sigma_{4}^{(0)})^{-1}.
\end{array}
\end{equation}
It follows from the condition (ii) of Theorem \ref{thm:1} and Lemma \ref{l:z*_0,n} that
$0<d/2\le |z^*_{0,n}-h_j^{(n)}|\le C$. Hence, (\ref{pr_ReS}) yields
\begin{equation}\label{vt_pr}
0<C_1<\dfrac{d^2}{d\,x^2}S_n(z^*_{0,n},\lambda_{0,n})\le C_2.
\end{equation}
We obtain
\begin{equation}\label{ReS_okr1}
\Re S_n(z,\lambda_{0,n})=\dfrac{d^2}{d\,x^2}S_n(z^*_{0,n},\lambda_{0,n})\dfrac{(x-z^*_{0,n})^2}{2}+
O((x-z^*_{0,n})^3).
\end{equation}
We get from (\ref{s1})
\[
s(x)\le 2|x-z^*_{0,n}|.
\]
This and the inequality $(x-z^*_{0,n})^2+s(x)\ge\delta^2$ imply for $x=\Re z$, $z\in L_n$, $|z-z^*_{0,n}|\ge\delta$
\begin{equation}\label{Re}
\delta^2/3\le |x-z^*_{0,n}|\le\delta.
\end{equation}
Thus, (\ref{vt_pr}) -- (\ref{ReS_okr1}) and the monotonicity of $\Re S_n(z_n(\lambda),\lambda_{0,n})$ for $\lambda>\lambda_{0,n}$
and $\lambda<\lambda_{0,n}$ (see Lemma \ref{l:min_max}) imply
\begin{equation}
\Re S_n(v,\lambda_0)\ge C\delta^4,\quad v\in L_n:\,\,|v-z^*_{0,n}|\ge\delta.
\end{equation}
We have proved the first inequality of Lemma \ref{l:vne_okr}.

To prove the second inequality consider $\Re S_n(z,\lambda_{0,n})$ for $z\in l_n$ of (\ref{l_n})
\begin{multline*}
\Re S_n(z^*_{0,n}+iy,\lambda_{0,n})=\dfrac{(z^*_{0,n})^2-y^2}{2}\\+\dfrac{1}{2n}\sum\limits_{j=1}^n\log
((z^*_{0,n}-h_j^{(n)})^2+y^2)-\lambda_{0,n}z^*_{0,n}-S^*.
\end{multline*}
Using (\ref{cond}) we get
\begin{multline*}
S_n(z^*_{0,n},\lambda_{0,n})=\dfrac{d}{d\,y}S_n(z^*_{0,n},\lambda_{0,n})\\
=\dfrac{d^2}{d\,y^2}S_n(z^*_{0,n},\lambda_{0,n})=
\dfrac{d^3}{d\,y^3}S_n(z^*_{0,n},\lambda_{0,n})=0
\end{multline*}
and
\[
\dfrac{d^4}{d\,y^4}S_n(z^*_{0,n},\lambda_{0,n})=-6\sigma_4^{(0)}.
\]
Since from condition (ii) of Theorem \ref{thm:1} and Lemma \ref{l:z*_0,n} we have
$0<d/2\le |z^*_{0,n}-h_j^{(n)}|\le C$, we obtain
\begin{equation}\label{ineq_pr_y}
-C<\dfrac{d^4}{d\,y^4}S_n(z^*_{0,n},\lambda_{0,n})<-c<0.
\end{equation}
Moreover,
\begin{equation*}
\Re S_n(z,\lambda_{0,n})=\dfrac{d^4}{d\,y^4}S_n(z^*_{0,n},\lambda_{0,n})y^4/4!+
O(y^5).
\end{equation*}
This, (\ref{ineq_pr_y}) and the monotonicity of $\Re S_n(z_{0,n}^*+iy,\lambda_{0,n})$ for $y>0$ and $y<0$
(see Lemma \ref{l:min_max}) imply
\begin{equation*}
\Re S_n(t,\lambda_0)\le -C\delta^4,\quad t\in l_n:\,\,|t-z^*_{0,n}|\ge\delta.
\end{equation*}
$\quad\Box$

\medskip

\textbf{Proof of Lemma \ref{l:as_ker}.}

Since $\lambda\not\in \hbox{supp}\,N$, it follows from the result of \cite{TSh:09} for a sufficiently small~$\delta$
\[
\displaystyle\int\dfrac{N^{(0)}(d\,h)}{(h-z(\lambda))^2}\le 1,\quad \lambda\in U_\delta(\lambda),
\]
where $z(\lambda)$ is defined in (\ref{z_lam}), and we have
\[
\lim\limits_{\varepsilon\to 0}\displaystyle\int\dfrac{\varepsilon
N^{(0)}(d\,h)}{|h-z(\lambda)-i\varepsilon|^2}=0,\quad\lambda\in U_\delta(\lambda).
\]
According to the Stieltjes-Perron formula, $N^{(0)}(z(U_\delta(\lambda)))=0$,
 hence $z(\lambda)\not\in \hbox{supp}\,N^{(0)}$.
using the same arguments as in Lemma \ref{l:z*_0,n}, we can prove that equation (\ref{eqv_f_0,n})
has only one root $z_n(\lambda)$ in $\omega=\{z\in\mathbb{C}: |z-z(\lambda)|\le \delta_1\}$, and
equation (\ref{eqv_f,pr}) does not have roots in $\omega$. Thus,
\begin{equation}\label{dist}
\hbox{dist}\,\{z_n(\lambda), L_n\}\ge C>0.
\end{equation}
Take $\widetilde{l}=\{z\in \mathbb{C}:z=z_n(\lambda)+iy,\,y\in \mathbb{R}\}$, move
integration in (\ref{K}) from $l$ to $\widetilde{l}$ and choose $L$ as $L_n$. We obtain
\begin{equation}\label{K1}
K_n(\lambda,\lambda)
=-n\displaystyle\int\limits_{\widetilde{l}}\displaystyle\frac{d\,t}{2\pi}\oint\limits_{L_n}
\displaystyle\frac{d\,v}{2\pi}
\displaystyle\frac{\exp\left\{ n(\widetilde{S}_n(t,\lambda)-\widetilde{S}_n(v,\lambda)))\right\}
}{v-t},
\end{equation}
where
\[
\widetilde{S}_n(z,\lambda)=z^2/2+\dfrac{1}{n}\sum\limits_{j=1}^n\log(z-h_j^{(n)})-\lambda z-
\widetilde{S}
\]
with $\widetilde{S}$ such that $\Re \widetilde{S}_n(z_n(\lambda),\lambda)=0$.
Similarly to Lemmas \ref{l:min_max} and \ref{l:vne_okr} we get for $t\in \widetilde{l}$, $v\in L_n$
\[
\Re \widetilde{S}_n(v,\lambda)\le -C<0,\quad
\Re \widetilde{S}_n(t,\lambda)\ge 0.
\]
This, (\ref{K1}), (\ref{dist}) and Proposition \ref{l:dl_kont} give the first assertion of Lemma \ref{l:as_ker}.
Moreover, according to (\ref{cond}) we get for $v\in L_n$
\[
\hbox{dist}(v,\{h_j^{(n)}\}_{j=1}^n)\le 1.
\]
Hence, the contour $L_n$ is bounded uniformly in $n$, and since for $t\in \widetilde{l}$ we have
$$\Re t=z_n(\lambda)=\lambda-1/\lambda+O(1/\lambda^2),\quad \lambda\to\infty,$$ we obtain
\[
\Re(\widetilde{S}_n(t,\lambda)-\widetilde{S}_n(v,\lambda))=
\dfrac{t^2-v^2}{2}+\dfrac{1}{n}\sum\limits_{j=1}^n\log\dfrac{t-h_j^{(n)}}{v-h_j^{(n)}}-
\lambda(t-v)\le-\dfrac{\lambda^2}{4}.
\]
This, (\ref{ots_besk}) and Lemma \ref{l:dl_kont} give (\ref{big_lam}).
$\quad\Box$

\section{Proof of Theorem \protect\ref{thm:2}}
Choose a sufficiently small $\delta>0$ and set
\begin{eqnarray}\label{Om_n}
\Omega_n=\left\{\{h_j^{(n)}\}_{j=1}^n \right.: &&\left|g_n^{(0)}(z)-f^{(0)}(z)\right|\le \dfrac{1}{n^{2/3+\alpha}},
\quad |z-z_0|\le\delta/2; \\ \notag
&& \left.\forall j=1,\dots,n \quad\hbox{dist}\,(h_j^{(n)},\hbox{supp}\,N^{(0)})\le \delta/10\right\}.
\end{eqnarray}
In the notation of Theorem \ref{thm:2} we have
\begin{multline*}
\mathbf{E}_n\bigg\{\prod\limits_{j=1}^n\bigg(1-\varphi\Big((n\gamma)^{2/3}(\lambda_j^{(n)}-\lambda_0)\Big)\bigg)\bigg\}\\
=\mathbf{E}^{(h)}_n\bigg\{(\mathbf{1}_{\Omega_n}+\mathbf{1}_{\Omega_n^C})
\mathbf{E}^{(g)}_n\bigg\{\prod\limits_{j=1}^n\bigg(1-\varphi\Big((n\gamma)^{2/3}(\lambda_j^{(n)}-\lambda_0)\Big)
\bigg)\bigg\}\bigg\},
\end{multline*}
where $\mathbf{E}^{(h)}_n$ and $\mathbf{E}^{(g)}_n$ are the expectation with respect the probability
law $\mathbf{P}^{(h)}_n$ of $H_n^{(0)}$ and $\mathbf{P}^{(g)}_n$ of $M_n$ of (\ref{M_n}) respectively.

Since $0\le\varphi(x)\le 1$, we get using conditions (i) -- (iii) of Theorem \ref{thm:2}
\begin{multline*}
\mathbf{E}^{(h)}_n\bigg\{\mathbf{1}_{\Omega_n^C}
\mathbf{E}^{(g)}_n\bigg\{\prod\limits_{j=1}^n\bigg(1-\varphi\Big((n\gamma)^{2/3}(\lambda_j^{(n)}-\lambda_0)\Big)
\bigg)\bigg\}\bigg\}\\ \le
\mathbf{E}^{(h)}_n\big\{\mathbf{1}_{\Omega_n^C}\big\}\to 0, \,\,n\to\infty.
\end{multline*}
We have to consider
\begin{equation*}
\mathbf{E}^{(h)}_n\bigg\{\mathbf{1}_{\Omega_n}
\mathbf{E}^{(g)}_n\bigg\{\prod\limits_{j=1}^n\bigg(1-\varphi\Big((n\gamma)^{2/3}(\lambda_j^{(n)}-\lambda_0)
\Big)\bigg)\bigg\}\bigg\}.
\end{equation*}
We have from the determinant formulas
\begin{multline}\label{gap_sl}
\mathbf{E}^{(h)}_n\bigg\{\mathbf{1}_{\Omega_n}
\mathbf{E}^{(g)}_n\bigg\{\prod\limits_{j=1}^n\bigg(1-\varphi\Big((n\gamma)^{2/3}(\lambda_j^{(n)}-\lambda_0)
\Big)\bigg)\bigg\}\bigg\}=\\
\mathbf{E}^{(h)}_n\bigg\{\mathbf{1}_{\Omega_n}\bigg(1+
\sum\limits_{l=1}^\infty\dfrac{(-1)^l}{l!}
\displaystyle\int\det\left\{\frac{1}{\gamma^{2/3}}
\mathcal{K}_n\left(\frac{x_i}{\gamma^{2/3}},
\frac{x_j}{\gamma^{2/3}}\right)\right\}_{i,j=1}^l\\
\times\prod\limits_{s=1}^l\varphi(x_s)
\prod\limits_{r=1}^ld\,x_r\bigg)\bigg\}.
\end{multline}
Since $\{h_j^{(n)}\}_{j=1}^n\in \Omega_n$ satisfy conditions of Theorem \ref{thm:1}, we conclude that the
r.h.s. of (\ref{gap_sl}) can be written as
\begin{multline*}
E_n^{(h)}\left\{\mathbf{1}_{\Omega_n}\left(\det(1-\varphi^{1/2}A\varphi^{1/2})+o(1)\right)\right\}\\=
\det(1-\varphi^{1/2}A\varphi^{1/2})+o(1), \,\, n\to\infty.
\end{multline*}
Theorem \ref{thm:2} is proved.

\textbf{Acknowledgements}
The author  is grateful to Prof. L.Pastur
 for many interesting discussion of the problem.

\end{document}